\documentclass[a4paper,11pt]{article}

\usepackage{geometry}
\usepackage{amsmath}
\usepackage{amssymb}
\usepackage{amsthm}
\usepackage{amsfonts}

\usepackage{graphicx}

\title{New Set of Codes for the Maximum-Likelihood \\
  Decoding Problem}
\author{M. Barbier}
\date{}

\newtheorem{Def}{Definition}[section]
\newtheorem{The}{Theorem}[section]
\newtheorem{Pro}{Proposition}[section]
\newtheorem{Rem}{Remark}[section]

\begin{document}

\maketitle

\begin{abstract}
  The maximum-likelihood decoding problem is known to be
  NP-hard for general linear and Reed-Solomon codes
  \cite{MLDProblemGeneral,MLDProblemRS}. In this paper, we introduce
  the notion of $\mathcal{A}$-covered codes, that is, codes that can
  be decoded through a polynomial time algorithm $\mathcal{A}$ whose
  decoding bound is beyond the covering radius. For these codes, we
  show that the maximum-likelihood decoding problem is reachable in
  polynomial time in the code parameters. Focusing on binary BCH
  codes, we were able to find several examples of
  $\mathcal{A}$-covered codes, including two codes for which the
  maximum-likelihood decoding problem can be solved in quasi-quadratic
  time.\\


\end{abstract}


\vspace{1cm}

\section{Introduction}

Berlekamp, McEliece and Van Tilborg showed in \cite{MLDProblemGeneral}
that the maximum-likelihood decoding is a NP-hard problem for
general linear codes. Guruswami and Vardy later proved in \cite{MLDProblemRS}
that this problem applied to the family of Reed-Solomon codes is also
NP-hard. We briefly recall below the maximum-likelihood problem.\\

\begin{Def}[Maximum-likelihood decoding problem]
  \label{Def:MLDP}
  Let $\mathcal{C}$ a linear code over $\mathbb{F}_q$ and 
  $v$ a $\mathbb{F}_q$-vector in the ambient space.
  The \em maximum-likelihood decoding problem \em is to find the
  codeword $w \in \mathcal{C}$ closest to $v$. Most precisely, to find
  $w \in \mathcal{C}$, such as 

  $$
  d(w,v) = d(v,\mathcal{C}) = \min_{c \in \mathcal{C}} \lbrace
  d(v,c)\rbrace.\\
  $$
\end{Def}

Clearly, if for a given code there exists an algorithm able to correct
a number of errors at least equal to the covering radius, then this
algorithm solves the maximum-likelihood decoding problem. We recall
the covering radius definition, which is the largest distance between
any vector in ambient space and the code.\\

\begin{Def}[Covering radius]
  \label{Def:CR}
  Let $\mathcal{C}$ a linear code over $\mathbb{F}_q$. Its ambient
  space is a $\mathbb{F}_q$-vector space $V$. Let $v\in V$, the \em
  covering radius \em $R$ of $\mathcal{C}$ is given by
  $$
  R = \max_{v \in V}\lbrace \min_{c \in \mathcal{C}}d(v,c)\rbrace.\\
  $$
\end{Def}

 In light of Wu's recent algorithmic advances in list decoding
 \cite{Wu}, we proceed in a comparaison between covering radii and
 now achievable decoding bounds with such algorithm. This leads us to
 propose the new algorithmic notion of {\em $\mathcal{A}$-covered
   codes} for which maximum-likelihood decoding problem can be carried
 out in polynomial time, and provide some examples by focussing the
 family of binary BCH codes. We also exhibit two codes for
 which the maximum-likelihood decoding problem has quasi-quadratic
 complexity.\\

\section{$\mathcal{A}$-covered codes}

In the rest of this paper, we follow the standard notations of 
\cite{CoveringCodes} and shall denote by $R$ the covering radius of a
code $\mathcal{C}$, and by $t\triangleq \lfloor \frac{d-1}{2}\rfloor$
its error correction capacity. We now recall the definition of a
perfect code.\\

\begin{Def}[Perfect code]
  A code $\mathcal{C}$ with capacity $t$ and covering radius $R$ is
  called a \em perfect code \em if and only if
  $$
  R = t.\\
  $$
\end{Def}

These codes are of course very interesting from a decoding point of
view since each element of their ambient spaces can be
decoded. Linear perfect codes are completely classified and for each
of them, we know a decoding algorithm up to $t = R$. The
maximum-likelihood problem is consequently trivial for perfect
codes. This very property prompts us to propose the notion of
\textit{$\mathcal{A}$-covered codes} in the context of list
decoding. We first introduce the following definitions:

\begin{Def}[List decoding algorithm]
  Let $\mathcal{C}$ a code and $v$ a word in its ambient
  space. $\mathcal{A}$ is a \em list decoding algorithm for
  $\mathcal{C}$ \em up to $\tau_{\mathcal{A}}$ if and only if it
  returns all codewords $w\in \mathcal{C}$ such that $d(v,w) \le
  \tau_{\mathcal{A}}$.\\ 
\end{Def}

\begin{Def}[Polynomial time list decoding algorithm]
  Let $\mathcal{C}$ a code, $n$ its length, $v$ a word in its ambient
  space and $\mathcal{A}$ a list decoding algorithm up to
  $\tau_{\mathcal{A}}$. $\mathcal{A}$ is a \em polynomial time
  list decoding algorithm \em if it runs in
  $\mathcal{O}(f(n))$, where $f(X) \in \mathbb{R}[X]$.\\
\end{Def}

We can now present the notion of \em $\mathcal{A}$-covered code\em.

\begin{Def}[$\mathcal{A}$-covered code]
   Let $\mathcal{C}$ a code with covering radius $R$ and $\mathcal{A}$ a
   polynomial time list decoding algorithm which decodes
   $\mathcal{C}$ up to $\tau_{\mathcal{A}}$. $\mathcal{C}$ is an \em
   $\mathcal{A}$-covered code \em if and only if
   $$
   R \le \tau_{\mathcal{A}}.\\
   $$
\end{Def}

\begin{Rem}
  Since this algorithm runs in polynomial time, the returned list is
  also of polynomial size.\\
\end{Rem}

\begin{Pro}
  \label{Prop:APC}
  Let $\mathcal{C}$ an \em $\mathcal{A}$-covered code\em. The
  maximum-likelihood decoding problem for $\mathcal{C}$, (as given by
  Definition \ref{Def:MLDP}) is solvable in a time polynomial in the
  code parameters.\\
\end{Pro}




As seen before, the notion of \textit{$\mathcal{A}$-covered code} can be
seen as a computational analogue to perfect codes, albeit in the list
decoding context (see Figure \ref{Fig:PCAPC}).\\

\begin{figure}[h]
  \begin{center}
    \begin{tabular}{c|c}
      Unique decoding & List decoding \\
      
      \hline

      &\\
      Perfect code & $\mathcal{A}$-covered code \\
      & \\

      $R = t$ & $R \le \tau_{\mathcal{A}}$\\

    \end{tabular}
  
    \caption{
      \label{Fig:PCAPC}
      \textit{Perfect code vs $\mathcal{A}$-covered code}}
  \end{center}
\end{figure}

\section{Case of binary BCH codes}

\begin{figure}[h]
  \begin{center}
    \includegraphics{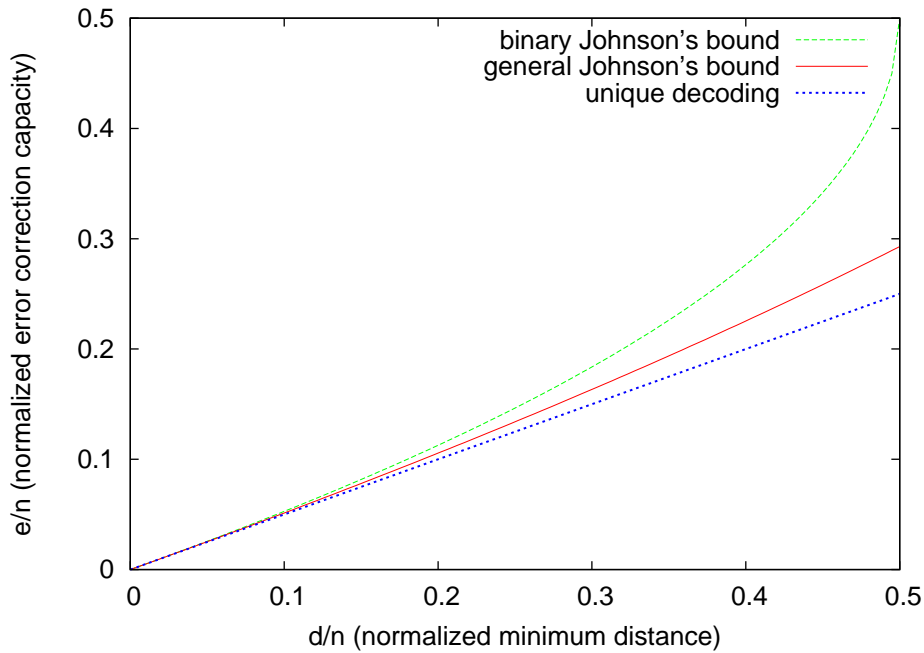}
    \caption{\label{Fig:Johnson_n}
      \textit{General and binary Johnson's bound}}
  \end{center}
\end{figure}

While still relatively recent, Wu's list decoding algorithm \cite{Wu}
is already regarded as a significant advance in the coding community.
Compared to the Guruswami-Sudan algorithm \cite{Guruswami}, it
exhibits an even better complexity. Moreover, when restricted to
binary BCH codes, Wu's method allows decoding up to the binary Johnson
bound $\frac{n}{2}(1-\sqrt{1-\frac{2d}{n}})$, whereas Guruswami-Sudan
only reaches the smaller general Johnson bound
$n(1-\sqrt{1-\frac{d}{n}})$, as shown in Figure \ref{Fig:Johnson_n}.\\

It is well known that $1$-error-correcting BCH codes are perfect
(these are Hamming codes) and $2$-error-correcting codes are
quasi-perfect \cite{CoveringCodes}. Since Wu's method is a polynomial
time list decoding algorithm \cite{Wu}, it is natural to compare their
covering radii and the binary Johnson bound of other binary BCH
codes. Unfortunately, classifying codes is a hard problem since it
requires to compute the covering radii which usually is not an easy
task \cite{CoveringRadiusPB}. Putting together and completing data
from the literature \cite{CoveringCodes}, we still manage to obtain
the list in Table \ref{Tab:comparaison}. This table includes all
primitive binary BCH codes of known covering radius. The
non-primitive, Wu-covered binary BCH codes of length 17 and 23 were
obtained by our own calculations.\\

\begin{table}[h]
  $$
  \begin{array}{|c|c|c|c||c|c|}
    \hline
    n & k & d & R  & \tau & \text{Comments}\\
    \hline
    
    \hline
    7 & 4 & 3 & 1 & 2 & \text{Hamming}\\
    
    \hline
    15 & 11 & 3 & 1 & 1 & \text{Hamming}\\

    15 & 7 & 5 & 3 & 3 & \text{Wu-covered code}\\

    15 & 5 & 7 & 5 & 5 & \text{RM(1,4)}^*\\

    \hline

    17 & 9 & 5 & 3 & 3 & \text{Wu-covered code}\\

    \hline

    23 & 12 & 7 & 3 & 4 & \text{Wu-covered code}\\
    
    \hline
    31 & 26 & 3  & 1  & 1 & \text{Hamming}\\
    31 & 21 & 5  & 3  & 2 & \\
    31 & 16 & 7  & 5  & 4 & \\
    31 & 11 & 11 & 7  & 7 & \text{Wu-covered code}\\
    31 & 6  & 15 & 11 & 12 & \text{RM(1,5)}^*\\

    \hline
    63 & 57 & 3  & 1 & 1 & \text{Hamming}\\
    63 & 51 & 5  & 3 & 2 & \\
    63 & 45 & 7  & 5 & 3 & \\
    63 & 39 & 9  & 7 & 4 & \\
    63 & 36 & 11 & 9 & 6 & \\

    \hline
  \end{array}
  $$
  \caption{\label{Tab:comparaison}
    \textit{Table of covering radius and binary
    Johnson bound for some binary BCH codes.
    }
  }
  
  \hspace{1cm}
\end{table}

Note that the BCH codes having 3 as their minimum distance are
Hamming codes. Since those are perfect codes, the maximum-likelihood
decoding problem is trivial. We also found two Reed-Muller codes of
first order \cite{Sloane}. Since the dimensions of first order
Reed-Muller codes are equal to the logarithm of their lengths, naively
listing all closest codewords is already a polynomial time decoding
algorithm. Hence, knowing that these codes are {\em Wu-covered} is of
little pratical important in solving the maximum-likelihood decoding
problem, since easier polynomial time methods are already available.\\

By contrast, the four codes $[15,7,5]$, $[17,9,5]$, $[23,12,7]$ and
$[31,11,11]$ given in Table \ref{Tab:comparaison} do not fall into
the two aforementioned families and we would expect the
maximum-likelihood decoding problem to be asymptotically
hard. However, the fact that they are {\em Wu-covered} implies that
this problem is actually solvable in polynomial time only (in the code
parameters).\\

\section{Quasi-quadratic list decoding of some binary BCH codes}

Guruswami-Sudan's algorithm can decode up to Johnson's bound in
polynomial time. As McEliece remarked in \cite{McEliece}, if we accept
to decode slightly less than this bound, the algorithm complexity is
dramatically reduced. Under this relaxed constraint, Wu demonstrated in
\cite{Wu} that his algorithm runs in quasi-quadratic time.\\


\begin{The}
  Wu's list decoding algorithm decodes up to 
  $$
  \tau = \lfloor \epsilon t +
  (1-\epsilon)\frac{n-n\sqrt{1-\frac{2d}{n}}}{2}\rfloor,
  $$
  with multiplicity $m = \lfloor \epsilon^{-1}\rfloor$
  in $\mathcal{O}(n^2 \lfloor \frac{1}{\epsilon} \rfloor^{4})$.\\
\end{The}


  


Consequently, binary BCH codes having binary Johnson bound strictly
greater than their covering radii, such as binary BCH $[31,6,15] =$
RM$(1,5)^*$ and BCH $[23,12,7]$, can be decoded in quasi-quadratic time
up to, and including, their covering radii.\\

\section{Conclusion}

Working purely from an algorithmic point of view, we proposed a new
set of codes, the \em $\mathcal{A}$-covered codes\em, for which we
showed that the maximum-likelihood decoding problem, known as NP-hard
in the general case, is solvable in polynomial time.\\

The main difficulty in finding such codes lies in the computation of
covering radii. However there may be quite a few of those codes as we
exhibited nine binary BCH codes which are \em Wu-covered codes\em, of
which, four constitute a new result and two can be decoded in time
quasi-quadratic in code parameters.\\

\bibliographystyle{plain}
\bibliography{biblio}

\end{document}